\documentclass[twocolumn, howpacs,superscriptaddress,article,floatfix]{revtex4}
\usepackage{graphicx}
\usepackage{amsmath}
\usepackage{natbib}
\usepackage{bm}
\usepackage[dvips]{color}
\begin{document}
\title{Ab initio prediction of magnetically dead layers in freestanding $\gamma$-Ce(111)}
\date{\today}
\author{S. Jalali Asadabadi}
\email[Electronic address:]{sjalali@phys.ui.ac.ir}
\author{F. Kheradmand}
\affiliation{Department of Physics, Faculty of Science, University
of Isfahan (UI), Hezar Gerib Avenue, Isfahan 81744, Iran}

\begin{abstract}
It is well known that the surface of nonmagnetic $\alpha$-Ce is
magnetically ordered, i.e., $\gamma$-like. One then might
conjecture, in agreement with previous theoretical predictions, that
the $\gamma$-Ce may also exhibit at its surfaces even more strongly
enhanced $\gamma$-like magnetic ordering. Nonetheless, our result
shows that the (111)-surfaces of magnetic $\gamma$-Ce are neither
spin nor orbitally polarized, i.e., $\alpha$-like. Therefore, we
predict, in contrast to the nonmagnetic $\alpha$-phase which tends
to produce magnetically ordered $\gamma$-like thin layers at its
free surfaces, the magnetic $\gamma$-phase has a tendency to form
$\alpha$-like dead layers. This study, which explains the suppressed
(promoted) surface magnetic moments of $\gamma$-Ce ($\alpha$-Ce),
shows that how nanoscale can reverse physical properties by going
from bulk to the surface in isostructural $\alpha$- and
$\gamma$-phases of cerium. We predict using our freestanding surface
results that a typical unreactive and non-diffusive substrate can
dramatically influence the magnetic surface of cerium thin films in
contrast to most of the uncorrelated thin films and strongly
correlated transition metals. Our result implies that magnetic
surface moments of $\alpha$-Ce(111) can be suddenly disappeared by
increasing lattice mismatch at the interface of a typical unreactive
and non-diffusive substrate with cerium overlayers.

\end{abstract}
 \maketitle

\section{Introduction} \label{sec-Int}
Cerium even in its bulk form has been a long-standing challengeable
system for theorists\cite{Eri92, Wan08} and experimentalists
\cite{Hom87, Ded08} due to its demonstrated variable degrees of 4f
states correlations,\cite{Pri99, Lig04, Ama08} resulting in various
astonishing properties.\cite{Ols85, Van01} Ce has been frequently
appraised to present or criticize strongly correlated
theories.\cite{Pau47, Mot68, Joh74, All82, Eri91, Ani91, Eli98,
Ani97, Lic98, Nik99, Hau05, Kot06} The reliability of the
theoretical approaches has not yet been fully authenticated for this
simple lanthanide metal. \cite{Mcm03} Cerium displays an intriguing
complex phase diagram containing a variety of solid
states.\cite{Ram71} In this diagram there are two much more
prominent phases, i.e., isostructural $\alpha$ and $\gamma$ solid
states. A normal Curie-Weiss law describes the magnetic
susceptibility of the $\gamma$-phase, while the $\alpha$-phase shows
an enhanced Pauli-like behavior.\cite{Ols85, Mur93} The $\gamma$-Ce
is a ferromagnetic system with localized 4f states, whereas the
$\alpha$-Ce is a paramagnetic system and its 4f states tend to be
more hybridized with the valence bands.\cite{Ama06} The $\gamma$-Ce
 crystallizes in the fcc structure with
$\sim15\%$ larger volume than its isostructural $\alpha$-Ce phase.
The isostructural $\alpha\longleftrightarrow\gamma$ phase transition
has been intensively studied.\cite{Mcm03, All82, Lae99, Ama06}
However, the physics underlying this unique transition is still
under debate.\cite{Ama06} The difference between $\alpha$- and
$\gamma$-phases is attributed to dissimilarity in magnitude of the
hybridization between 4f$^1$ and 5d$^1$6s$^2$ states.\cite{Ama08}
Now let us switch to its surface states, which can be even more
problematic than the bulk states. The surface electronic states of
cerium can be different from that of its bulk due to the appearance
of surface states\cite{Wes98} and sensitivity of cerium to the
symmetry of its surrounding environment.\cite{Eri91} Eriksson et
al.\cite{Eri91}, within their linearized muffin-tin orbitals (LMTO)
calculations employing the Vosko-Wilk-Nusair local density
approximation (VWN-LDA), found $\alpha$-Ce(100) surfaces to be
$\gamma$-like, i.e., magnetically ordered. They\cite{Eri91} from
their later result anticipated that the surface magnetism at the
$\gamma$-Ce(100) layers should be more enhanced $\gamma$-like than
 the $\alpha$-Ce(100). The later anticipation might be a
natural consequence of the Stoner's picture that, "the tendency
toward magnetism should be increased near metal surfaces, because of
the narrowing of the density of states that yields a Stoner
enhancement in the susceptibility" \cite{Sto03}. Rothman and
coauthors\cite{Rot99} experimentally by growing Ce(111) on a W(110)
buffer confirmed the photo-emission spectra measurements of the Gu
et al.\cite{Gu91} in which $\alpha$-like spectra were observed for a
lattice parameter close to that of the $\gamma$-Ce. The later
observations\cite{Rot99, Gu91} yield a clue to think about the
possibility of vanishing magnetic moments at the surface of magnetic
materials, the so-called dead layers\cite{Sto03}. The validity of
the fact that the surface of $\alpha$-Ce(111) is $\gamma$-like has
extensively been experimentally\cite{Wes91, Wes98} and
theoretically\cite{Eri97, Pri99} verified. Nonetheless, there are
unfortunately few reports on the $\gamma$-Ce thin films to justify
about the magnetic tendency of the $\gamma$-Ce(111) surface to show
whether it would be $\gamma$-like or $\alpha$-like. These motivated
us to systematically calculate the surface magnetic moments of
$\alpha$- and $\gamma$-Ce(111) phases including spin and orbital
polarizations as well as spin-orbit coupling which can be compulsory
for cerium based systems.\cite{Jal07} We, in agreement with
Ref.~\onlinecite{Eri91} and experiment\cite{Wes91, Wes98}, show that
the surface of nonmagnetic $\alpha$-phase tends to behave as a
magnetic $\gamma$-like Ce compound. However, our result unexpectedly
implies that the surface of $\gamma$-Ce(111) is $\alpha$-like.
Indeed, our calculations demonstrate that the surface of magnetic
$\gamma$-Ce is neither spin- nor orbital-polarized. Our result
implies that an unreactive and non-diffusive substrate can cause a
magnetic transition from magnetically ordered layers to a
nonmagnetic layers at the surface of $\alpha$-Ce(111) thin films. We
shall discuss how cerium can cause such an unexpected magnetic
transition.

\section{Computational Details} \label{sec-Com}
This work has been carried out using the program package
WIEN2k,\cite{Bla01} which allows to perform accurate all-electron
full-potential augmented plane waves plus local orbital (APW+lo)
\cite{Sjs00, Mad01} band structure calculations of solids within the
density functional theory (DFT).\cite{Hoh64,Koh65} The
Perdew-Burke-Ernzerhof generalized gradient approximation
(PBE-GGA)\cite{Per96} has been used for the exchange-correlation
functional. The Ce Muffin-tin radii were set to $R_{MT}=1.8~a.u.$
for the surface calculations. The expansion of the wave functions
and charge densities were cut off by the $R_{MT}K_{max} = 7.0$ and
$G_{max} = 12$ parameters, respectively. A set of
$21\times21\times1$ ($19\times19\times19$) special k-points has been
used for integrations over the Brillouin zone of the supercell (unit
cell) in the surface (bulk) calculations. The full relaxations were
performed with the criterion of 1 mRy/bohr on the exerted forces.
The relativistic effects were taken into account by including the
spin-orbit (SO) coupling in a second variational procedure. Orbital
polarizations were included to consider appropriate correlation in
4f Ce states using the LDA+U method.\cite{Ani93} Since LDA+U
calculations can result in different solutions depending on initial
conditions, care has been made to use an appropriate density matrix
to ensure that the result has not been trapped in a local minimum.

\section{Superlattice Surface Structure} \label{sec-super}
\begin{figure}[!t]
 \begin{center}
  \includegraphics[width=9cm,angle=0]{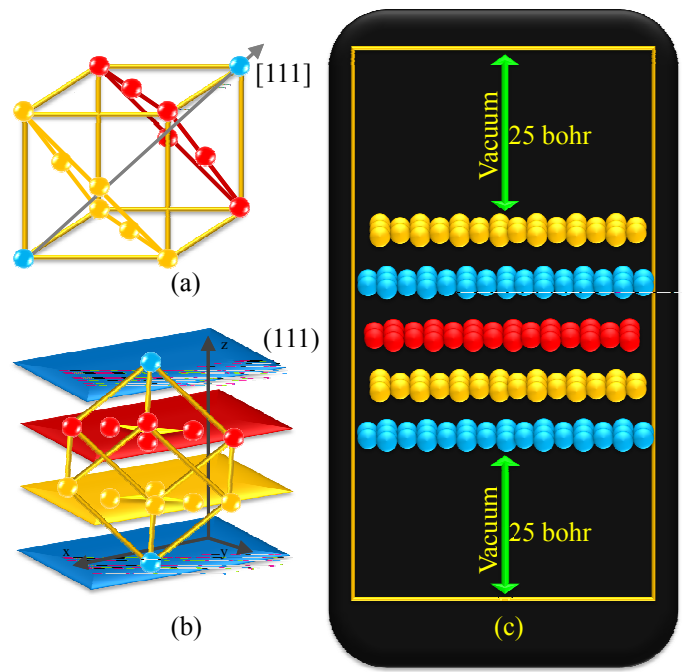}
  \caption{"(color online)" Visualization of slab construction: (a)
  Fcc unit cell, where atoms on different planes normal to the [111] direction are distinguished, (b) Diagonal view of the fcc unit cell
   along the [111] direction indicating (111) surfaces. (c) Supercell containing 5 Ce(111) layers immersed symmetrically in a vacuum space
    as our simulated slab.
 \label{fig1}}
 \end{center}
\end{figure}

Isostructural $\alpha$- and $\gamma$-Ce crystallize in the Fm\={3}m
space group symmetry with the face-centered cubic (fcc) lattice
structure, as shown in Fig.~\ref{fig1} (a). We illustrate in
Fig.~\ref{fig1} (b) a diagonal view of the fcc structure along the
[111] direction normal to the (111) planes. The z axis is chosen to
be perpendicular on the surfaces so that the x and y axes are
parallel to them, as shown in Fig.~\ref{fig1} (b). Thin-films can be
constructed by stacking cerium layers along our desired direction.
In order to simulate the free Ce(111) surfaces, a symmetric
supercell is created by adding vacuum spaces above and below Ce
layers, as shown in Fig.~\ref{fig1} (c). To determine the vacuum
thickness, total energies, work functions and exerted forces on the
surface atoms were calculated versus various vacuum thicknesses. Our
result shows that 25 bohr is sufficient for the vacuum thickness, as
illustrated in Fig.~\ref{fig1} (c), to avoid interactions with the
nearest neighbors of the slab along the Cartesian z axis.

\section{Bulk Properties} \label{sec-bulk}
In order to more reliably simulate the surfaces of the $\alpha$- and
$\gamma$-Ce(111) so that the surface properties of the
$\alpha$-phase can be distinguished from that of the $\gamma$-phase
in consistent with experiment, we would first investigate whether
the method used for the surface calculations can properly reproduce
the bulk properties of these two phases.  The ground state
equilibrium volumes were measured to be 28.17 $\AA^3$ for the
$\alpha$-Ce and 34.36 $\AA^3$ for the $\gamma$-Ce.\cite{Don74} The
most apparent difference between these two phases is that the
lattice parameter of the $\gamma$-phase is larger than that of the
$\alpha$-phase. To authenticate whether the degree of 4f
hybridization with other valence states can cause this change in
lattice parameter, we have applied the LDA+U method\cite{Ani93} to
the $\gamma$-phase and used the PBE-GGA\cite{Per96} for the
$\alpha$-phase. The spin-polarized PBE-GGA including spin-orbit
coupling (GGA+SP+SO) result shows an excellent agreement between our
calculated lattice parameter, 4.808 \AA, with experiment, 4.830
\AA,\cite{Don74} for the $\alpha$-phase. The later obtained
excellent agreement using solely the GGA+SP+SO, with no further band
correlated corrections, can be taken as an indication to the fact
that the 4f $\alpha$-Ce electrons tend to behave as band-like
itinerant electrons in agreement with Mott transition (MT)
scenario.\cite{Mot68, Joh74} The GGA+SP+SO plus correlations
(GGA+SP+SO+U) among 4f Ce electrons has been used with literature
values of 6.1 eV\cite{Ani91, Coc05} and 4.4 eV\cite{Coc05} for the
Hubbard U parameter in the $\gamma$-phase. We found that the former
U value, 6.1 eV, results in a completely wrong lattice parameter for
the $\gamma$-phase. For U = 0, the minimum of the E-V curve of the
localized $\gamma$-Ce coincides with the one of the $\alpha$-Ce. We
could calculate the lattice parameter of the $\gamma$-phase using
the later U value, which was corresponded\cite{Coc05} to the
selected atomic configurations of $U =
E(f^3S^0)+E(f^1S^2)-2E(f^2S^1) = 4.4~eV$ for the promoted localized
4f electrons to more delocalized 6s states. Our calculated lattice
parameter of $\gamma$-Ce, 5.169 \AA, using U = 4.4 eV is in
excellent agreement with experiment, 5.160 \AA, as well.\cite{Don74}
We obtained 11 kbar for the $\alpha\leftrightarrow\gamma$-phase
transition pressure which is comparable with experimental value of 8
kbar and other theoretical results.\cite{Pri99, All82, Sva96, Joh95}
Total magnetic moments were calculated to be 0.000 and 1.142 $\mu_B$
for the $\alpha$- and $\gamma$-phases, respectively, in agreement
with the SIC-LSD results.\cite{Sva96}

\section{Surface Properties} \label{sec-Sur}
\subsection{Work Function} \label{sec-Work}
Insofar as the accuracy of our surface calculations are concerned,
our success in reproducing bulk properties may not seem to be
sufficient. Thus, we would second offer an estimation concerning the
accuracy of our surface calculations by reporting on the work
function as an extremely surface sensitive quantity\cite{Bae91}.
Therefore, we have also calculated the work function ,$\phi (eV)$,
as the minimum energy required to liberate an electron from the
Fermi level ($E_F$) to a point with negligible kinetic energy at the
center of the vacuum of the slab. The work functions were obtained
using the relation of $\phi=E_{vac} - E_F$, where $E_{vac}$ is
estimated by the averaged electrostatic Coulomb potential at the
midpoint of the vacuum of the slab and $E_F$ is the corresponding
Fermi energy.

\begin{figure}[!t]
 \begin{center}
  \includegraphics[width=9cm,angle=0]{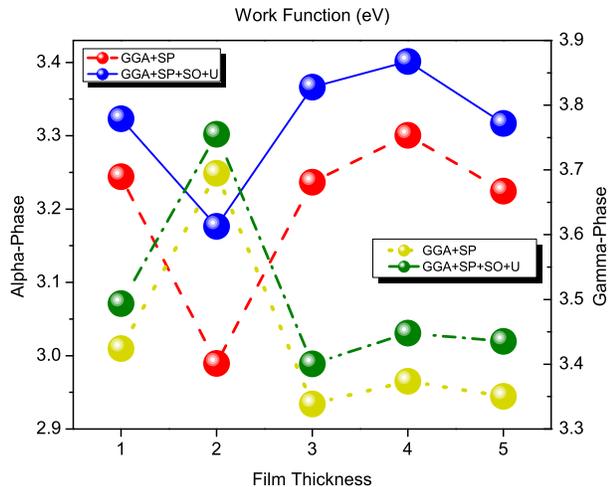}
  \caption{"(color online)" Work function, $\phi$ (eV), versus
  number of cerium layers for $\alpha$-Ce(111), left vertical axis, and
  $\gamma$-Ce(111), right vertical axis, within GGA+SP and GGA+SP+SO+U.
 \label{fig2}}
 \end{center}
\end{figure}

The work functions are given in Fig.~\ref{fig2} as a function of the
film thickness for both of the $\alpha$- and $\gamma$-C(111) clean
surfaces. As shown in Fig.~\ref{fig2}, the SO+U causes to reduce the
work function for both of the phases. The result shows that the work
functions of the $\gamma$-phase is well converged through our five
thin layers. The convergence of the $\gamma$-phase, as shown in
Fig.~\ref{fig2}, is more reliable than that of the $\alpha$-phase.
Fluctuation of the $\alpha$-phase between fourth and fifth layers
shows also a small change of about 2\% in the work functions. The
result, Fig.~\ref{fig2}, indicates that the fluctuation of the work
function is not substantially affected by the spin-orbit coupling
and/or LDA+U correlations. The result shows that the work function
oscillate as a function of the number of layers with the period of
$\lambda = 2$ layers. The oscillation indicates how the quantum size
effects (QSEs) \cite{Raf09} can affect the results. The period of
oscillation, as shown in Fig.~\ref{fig2}, is less than the period of
the "abcabc" pattern of the fcc stacking. This can be taken as an
indication to the fact that here oscillations of the work function
originates more dominantly from the quantum size effects (QSEs) at
nanoscale than the period of thin films stacking due to the symmetry
of the structure\cite{Raf09}. Our result shows that the work
functions of the $\gamma$-phase are larger than those of the
$\alpha$-phase, for every number of layers. The later result is in
agreement with the LMTO result,\cite{Eri91} where the work functions
were calculated to be 3.5 eV for $\alpha$-Ce(100) and 4.2 eV for
$\gamma$-Ce(100). Our calculated work functions are closer to the
experimental value, 2.9$\pm$0.2,\cite{Eas70}  than the LMTO values.
These results, in accord with the MT\cite{Mot68, Joh74} and Kondo
volume collapse (KVC)\cite{All82} pictures, which reconfirm that the
4f electrons tend to be more localized in the $\gamma$-Ce than that
of $\alpha$-Ce, may ensure that our subsequent surface results might
be reliable as well.

\subsection{Magnetic Moment} \label{sec-Mag}

\begin{table}[!t]
 \begin{center}
 \caption{Spin magnetic moments inside the Muffin-tin sphere (MT), spin
 magnetic moments in the interstitial region (Int.), total spin
 magnetic moments per atom (Spin), orbital magnetic moment inside the Muffin-tin
 sphere (Orb.) and total magnetic moments per atom (Tot.) in $\mu_B$ versus number of
 layers (N) within the GGA+SP+SO for the $\alpha$-Ce(111) thin  films.
 We calculated total
spin moment as Spin = MT+Int., for N = 1 and as Spin =
MT+$\frac{1}{2}$Int., for
 N$\geq$2. Total magnetic moment is calculated to be Tot. = Spin + Orb., for every N.
 \label{tab1}}
  \begin{ruledtabular}
   \begin{tabular}{lcccccc}
    N&1&2&3&4&5&\\
    \hline
    MT              &0.61&1.02&0.53&0.49&0.65&\\
    Int.            &0.37&1.54&0.85&0.71&0.75&\\
    Spin            &0.98&1.79&0.96&0.85&1.03&\\
    Orb.            &-0.21&-0.45&-0.22&-0.22&-0.39&\\
    Tot.            &0.77&1.34&0.74&0.63&0.64&\\
   \end{tabular}
  \end{ruledtabular}
 \end{center}
\end{table}

Turning to the goal of this paper, we first report on the magnetic
moments for the $\alpha-$Ce(111) thin films and then for the
$\gamma-$Ce(111) thin films. The magnetic moments are listed in
Tab.~\ref{tab1} for the $\alpha-$Ce(111) thin films. The result
shows that the direction of the total spin moments is opposite to
that of the orbital moments. However, total magnetic moments are
still considerable for the surface of $\alpha$-Ce. This implies that
the surface of the nonmagnetic $\alpha$-Ce is $\gamma$-like in
agreement with experiments\cite{Wes91, Wes98} and other theoretical
results \cite{Eri97, Pri99} as well as the pioneer work of Eriksson
and coworkers\cite{Eri91}. The result, as presented in
Tab.~\ref{tab1}, shows that the total magnetic moment changes by
only a tiny percentage of about 1\% on going from fourth to the
fifth layer. Here in fact, cancellation errors between spin and
orbital contributions cause such a tiny percentage in the total
magnetic moments. The surface of $\alpha$-Ce(111) is well known to
be $\gamma$-like experimentally\cite{Wes91, Wes98} and
theoretically\cite{Eri97, Pri99}. Therefore, the result presented in
Tab.~\ref{tab1} can qualitatively corroborate the idea that the
surface of $\alpha$-Ce(111) is magnetically ordered in agreement
with theory\cite{Eri97, Pri99} and experiment\cite{Wes91, Wes98}.
For the $\gamma$-Ce(111) thin films, we did not present similar
table, because our result shows that such a table contains nothing
more than zero moments. The accuracy to which we have calculated the
magnetic moments is $\pm$0.01 $\mu_B$ for the $\gamma$-Ce(111).
Thereby if we present a table for the magnetic surfaces of
$\gamma$-Ce(111), the table contains ignorable values of 0.00 $\pm$
0.01 $\mu_B$. Such ignorable values are obtained for all components
of the total magnetic moments, i.e. according to the abbreviations
given and defined in the caption of Tab.~\ref{tab1}, MT, Int., Spin,
Orb. and as a result Tot., from N = 1 to N = 5. Therefore in
essence, inversely the surface of the magnetic $\gamma$-Ce is found
within our ab initio DFT calculations to be $\alpha$-like. Thus, we
predict that the free surfaces of the $\gamma$-phase constitute dead
layers at zero temperature. Such a contradictory situation in these
isostructural phases concerning different magnetic behaviors of
their bulks when compared with their free (111) surfaces can be
considered as another physical property for cerium. Our result
confirms the experimental works preformed by Rothman and
coauthors\cite{Rot99} and Gu et al.\cite{Gu91} where $\alpha$-like
spectra were measured for a lattice parameter close to that of the
$\gamma$-Ce. This represents another success of the density
functional theory (DFT) in fundamentally predicting complicated
4f-electron nanosystems from first principles of quantum mechanics
without assuming any experimental data. It is to some extent hard to
expect that there can be found another case in nature similar to
cerium with two isostructural phases and opposite magnetic surfaces
compared to their corresponded bulk counterparts. This implies that
cerium due to its demonstrated capacity in reproducing two entirely
opposite magnetic surface states might be considered as a unique
case. Thereby, Ce once more serves as an interesting case to
appraise the validity of our fundamental understanding based on the
density functional theory (DFT).

\section{Effect of Lattice Mismatch} \label{sec-DOS}

From our hypothetical freestanding surface study one may also deduce
that a lattice mismatch at the interface with a typical unreactive
and non-diffusive substrate can play an important role on the
magnitude of the magnetic moments in Ce thin films. The sensitivity
of cerium thin films to the effects of lattice mismatch can be
inferred by the fact that the differences between $\alpha$- and
$\gamma$-phases may be resulted from the larger lattice parameter of
the $\gamma$-Ce than that of the $\alpha$-Ce. The unreactive and
non-diffusive substrate affects the lattice mismatch of the cerium
overlayers. The lattice mismatch at the interface of a typical
Ce(111)/substrate causes to change lattice parameter of the cerium
thin films. The change of lattice parameter in Ce can give rise to
transition from $\alpha$- to $\gamma$-phase. In uncorrelated thin
films the change of lattice mismatch to some extent may not
drastically change the electronic structures. This could be also the
case for strongly correlated materials. For instance, the effects of
gold and cooper substrates on the ferromagnetic transition metals
were refuted in leading to magnetically dead layers.\cite{Lie70}
Lattice mismatch on cerium is important not only due to its
4f-electrons, but also more importantly due to existing
isostructural $\alpha$- and $\gamma$-phases in the vicinity of each
other. This point might be of significant importance in
nanotechnology, as in this case one can control the surface magnetic
moments ranging from enhanced magnetically ordered surfaces to
magnetically dead layers by changing unreactive and non-diffusive
substrates. Therefore, we predict that more lattice mismatch in
$\alpha$-Ce thin films can give rise to less magnetic surface. The
magnetic surface of $\alpha$-Ce thin films can suddenly disappear by
increasing lattice mismatch towards lattice parameter of
$\gamma$-Ce. The later point shows that how an unreactive and
non-diffusive substrate can significantly influence the cerium thin
films.

\begin{figure} [!t]
 \begin{center}
  \includegraphics[width=9.0cm,angle=0]{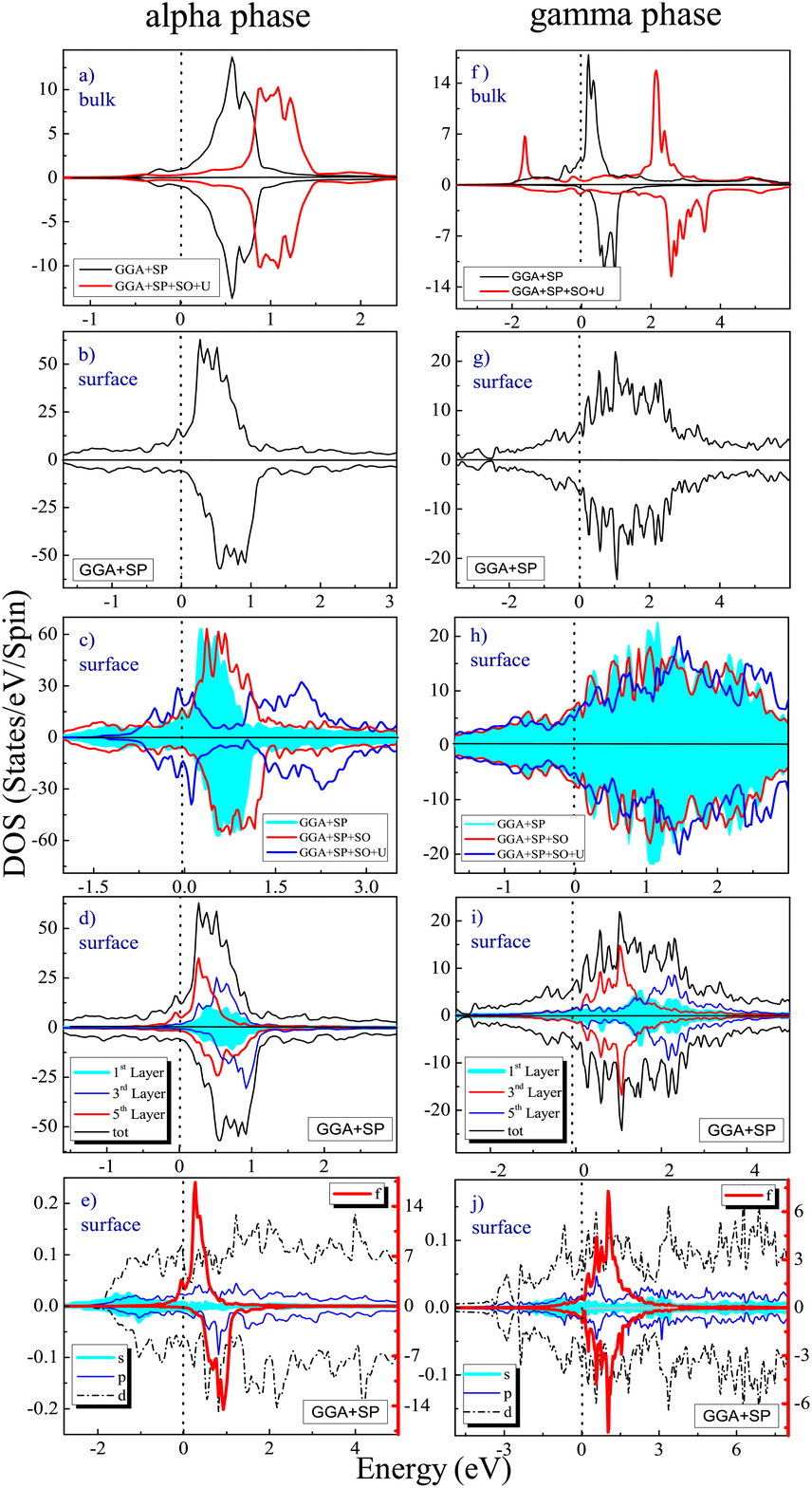}
  \caption{"(color online)" Bulk and free surface DOSs of the $\alpha$-Ce (left panel) and $\gamma$-Ce (right panel).
 \label{fig3}}
 \end{center}
\end{figure}

\section{Density of States (DOS)} \label{sec-DOS}
To realize the underlying physics of the above discussed property,
we give in Figs.~\ref{fig3} the density of states (DOS) curves. As
shown in Fig.~\ref{fig3} (a), the up and down bulk DOSs are
symmetric with respect to each other for the $\alpha$-Ce, which is
not the case for the $\gamma$-Ce, as shown in Fig.~\ref{fig3} (f).
Therefore, one anticipates to find (zero) nonzero magnetic moment
for the ($\alpha$-phase) $\gamma$-phase. As can be seen from
Figs.~\ref{fig3} (a) and (f), including correlations among
4f-electrons cannot affect the later anticipation. The calculated
surface DOS curves within LDA are shown in Figs.~\ref{fig3} (b) and
(g) for the $\alpha$- and $\gamma$-Ce(111), respectively. One
notices from the later surface DOSs that the $\alpha$- and
$\gamma$-phases would exchange their magnetic behaviors by going
from their bulks to their surfaces. As shown in Fig.~\ref{fig3} (g),
the up and down bulk DOSs are symmetric with respect to each other
for the $\gamma$-Ce(111) surface, which is not the case for the
$\alpha$-Ce(111), as shown in Fig.~\ref{fig3} (b). Therefore, in
contrast to what has been already anticipated above for the bulk
counterparts, at the surface the $\alpha$-Ce(111) is magnetic
whereas $\gamma$-Ce(111) is nonmagnetic. This shows that the above
mentioned anticipation can be affected by the surface states. This
means that the surface of the $\alpha$-Ce is $\gamma$- like whereas
the surface of the $\gamma$-Ce is $\alpha$-like. As can be seen from
Figs.~\ref{fig3} (c) and (h), including spin-orbit (SO) coupling and
correlations among 4f-electrons cannot change back the reversed
surface magnetic properties to that of the bulks. To shed light into
the electronic structures of interior layers, DOSs at various layers
of our slab containing totaly 5 symmetric layers were also
calculated within the GGA+SP+SO+U and the results are given in
Figs.~\ref{fig3} (d) and (i) for the $\alpha$- and $\gamma$-Ce(111),
respectively. The result shows that all the layers, i.e., deeper
and/or shallower layers, are magnetic for the $\alpha$-Ce(111) and
nonmagnetic for the $\gamma$-Ce(111). The calculated Partial DOSs
employing GGA+SP+SO+U are shown in Figs.~\ref{fig3} (e) and (j) for
the $\alpha$- and $\gamma$-Ce(111), respectively. The result, as can
be seen from Fig.~\ref{fig3} (e), shows that the surface magnetic
properties of the nonmagnetic $\alpha$-phase is due to the partial
4f-states. The result, as shown in Fig.~\ref{fig3} (j), also clarify
the role of 4f-states in suppressing the magnetic moments at the
surface of magnetic $\gamma$-phase.

\section{Conclusion} \label{sec-con}
In conclusion, we have presented an evidence in favor of entirely
different magnetic orderings at the $\alpha-$ and $\gamma-$Ce(111)
clean surfaces as opposed to their bulk counterparts. Bulk magnetic
moments were calculated to be zero and 1.14 $\mu_B$ for
isostructural enhanced Pauli-like $\alpha-$Ce and Curie-Weiss
$\gamma-$Ce phases, respectively, in agreement with
 other theoretical calculations. Inversely, surface
magnetic moments were calculated to be nonzero for the
$\alpha-$Ce(111) and zero for the $\gamma-$Ce(111) thin film layers.
The contradictory situation that arises from different magnetic
transitions in these isostructural phases from their bulks to their
free (111) surfaces can be considered as another unexpected property
for cerium and attributed to different 4f hybridizations with the
other valence states at diverse surface circumstances. Our result
predicts that the nonmagnetic $\alpha$-phase tends to produce
magnetically ordered $\gamma$-like surface layers, whereas the
magnetic $\gamma$-phase has a tendency to form $\alpha$-like dead
layers. In summery, we have shown that the surface of $\alpha$-Ce is
$\gamma$-like, while the surface of $\gamma$-Ce is $\alpha$-like.
Within the later result one concludes that how can cerium be of
significant importance in nanotechnology, because physical
properties of cerium thin films can be drastically influenced by
those of unreactive and non-diffusive substrates which can impose
large lattice mismatch. Our result implies that the significant
effects of unreactive and non-diffusive substrates on the adatoms
can be expected solely from cerium due to its 4f-electrons and
isostructural $\alpha$- and $\gamma$-phases as well as the discussed
entirely different magnetic ordering of the isostructural phases at
their corresponded surfaces.

\acknowledgments This work is supported by University of Isfahan
(UI), Isfahan, Iran. S.J.A. is thankful to Computational
Nanotechnology Supercomputing Center Institute for Research in
Fundamental Science (IPM) P.O.Box 19395-5531, Tehran, Iran for the
computing facility.


\begin{thebibliography}{}
\bibitem{Eri92} O. Eriksson, J.M. Wills, and A.M. Boring, Phys. Rev. B {\bf46}, 12981-12989 (1992).
\bibitem{Wan08} Y. Wang et al., Phys. Rev. B {\bf78}, 104113 (2008).
\bibitem{Hom87} Hitoshi Homma, Kai-Y. Yang, and Ivan K. Schuller, Phys. Rev. B {\bf36}, 9435-9438 (1987).
\bibitem{Ded08} Yu. S. Dedkov et al., Phys. Rev. B {\bf76}, 073104 (2007).
\bibitem{Pri99}D.L. Price, Phys. Rev. B {\bf60}, 10588 (1999).
\bibitem{Lig04} B.E. Light et al., Phys. Rev. B {\bf69}, 024419 (2004).
\bibitem{Ama08} B. Amadon, F. Jollet, and M. Torrent, Phys. Rev. B {\bf77},
155104 (2008).
\bibitem{Ols85} J.S. Olsen et al., Physica B $\&$ C {\bf133}, 129 (1985).
\bibitem{Van01}J.W. van der Eb, A.B. Ku\'{z}menko, and D. van der Marel, Phys. Rev.
Lett. {\bf86}, 3407 (2001).
\bibitem{Pau47}L. Pauling, J. Am. Chem. Soc. {\bf69}, 542 (1947).
\bibitem{Mot68} N.F. Mott, Rev. Mod. Phys. {\bf40}, 677 (1968); Metal-Insulator
Transitions (Taylor $\&$ Francis, London, 1990).
\bibitem{Joh74}B. Johansson, Philos. Mag. {\bf30}, 469 (1974).
\bibitem{All82}J.W. Allen and R.M. Martin, Phys. Rev. Lett. {\bf49}, 1106 (1982); M.
Lavagna, C. Lacroix, and M. Cyrot, Phys. Lett. {\bf90A}, 210 (1982);
J.W. Allen and L.Z. Liu, Phys. Rev. B {\bf46}, 5047 (1992).
\bibitem{Eri91} O. Eriksson et al., Phys. Rev. B {\bf43}, 3137-3142 (1991).
\bibitem{Ani91} V.I. Anisimov and O. Gunnarsson,
 Phys. Rev. B {\bf 43}, 7570 (1991).
\bibitem{Ani97}V.I. Anisimov et al., J. Phys.: Condens. Matter {\bf9}, 7359 (1997).
\bibitem{Lic98} A.I. Lichtenstein and M.I. Katsnelson, Phys. Rev. B {\bf57}, 6884 (1998).
\bibitem{Eli98}G. Eliashberg and H. Capellmann, Pi\'{s}ma Zh. Eksp. Teor. Fiz {\bf67}, 111
(1998); JETP Lett. {\bf67}, 125 (1998).
\bibitem{Nik99}A.V. Nikolaev and K.H. Michel, Eur. Phys. J. B {\bf9}, 619 (1999); {\bf17},
12 (2000); Phys. Rev. B {\bf66}, 054103 (2002).
\bibitem{Hau05} Kristjan Haule et al., Phys. Rev. Let. {\bf94}, 036401 (2005).
\bibitem{Kot06} G. Kotliar et al., Rev. Mod. Phys. {\bf78}, 865 (2006).
\bibitem{Mcm03}A.K. McMahan, K. Held, and R.T. Scalettar, Phys. Rev. B {\bf67}, 075108
(2003).
\bibitem{Ram71}R. Ramirez, and L.M. Falicov, Phys. Rev. B {\bf3}, 2425
(1971).
\bibitem{Mur93}A.P. Murani et al., Phys. Rev. B {\bf48}, 13981-13984
(1993).
\bibitem{Ama06}B. Amadon et al., Phys. Rev. Lett. {\bf96}, 066402
(2006).
\bibitem{Lae99} J. L{\ae}gsgaard and A. Svane, Phys. Rev. B {\bf59}, 3450-3459 (1999).
\bibitem{Wes98}E. Weschke et al., Phys. Rev. B {\bf58}, 3682 (1998).
\bibitem{Sto03} N. Stoji\'{c}, J.W. Davenport, M. Komelj, and J. Glimm, Phys. Rev. B {\bf68}, 094407 (2003).
\bibitem{Bae91}Y. Baer et al., Phys. Rev. B {\bf44}, 9108 (1991).
\bibitem{Wes91}E. Weschke et al., Phys. Rev. B {\bf44}, 8304 (1991).
\bibitem{Eri97}O. Eriksson et al., Surface Science {\bf382}, 93 (1997).
\bibitem{Rot99}J. Rothman et al., Jou. of Mag. and Mag. Mat., {\bf198}, 276 (1999).
 \bibitem{Gu91}C. Gu,  et al., Phys. Rev. lett. {\bf67}, 1622 (1991).
 \bibitem{Jal07}S. Jalali Asadabadi, Phys. Rev. B {\bf75}, 205130 (2007).
\bibitem{Bla01}P. Blaha et al., WIEN2K, "An Augmented Plane Waves + Local Orbitals Program
for Calculating Crystal Properties," Karlheinz Schwarz, Techn.
Universitat Wien, Austria, ISBN 3-9501031-1-2 (2001).
\bibitem{Sjs00} E. Sj\"{o}stedt, L. Nordstr\"{o}m,
 and D. J. Singh, Solid State Commun. {\bf 114}, 15 (2000).
\bibitem{Mad01} G.K.H. Madsen et al., Phys. Rev. B {\bf 64}, 195134 (2001).
\bibitem{Hoh64} P. Hohenberg and W. Kohn, Phys. Rev. {\bf 136}, 864 (1964).
\bibitem{Koh65}W. Kohn and L. J. Sham, Phys. Rev. {\bf 140}, A1133 (1965).
\bibitem{Per96} J.P. Perdew, K. Burke, and M. Ernzerhof,
 Phys. Rev. Lett. {\bf 77}, 3865 (1996).
\bibitem{Ani93}V.I. Anisimov et al., Phys. Rev. B {\bf 48}, 16929 (1993); M. T. Czy\.{z}yk and G. A. Sawatzky,
 Phys. Rev. B {\bf 49}, 14211 (1994); E. Sj\"{o}stedt, L. Nordstr\"{o}m,
 and D. J. Singh, Solid State Commun. {\bf 114}, 15 (2000).
\bibitem{Don74}J. Donohu, The Stucture of the Elements (John Wiley $\&$ Sons, Inc., New York, 1974).
\bibitem{Coc05}M. Cococcioni and S. de Gironcoli, Phys. Rev. B {\bf 71}, 035105 (2005).
\bibitem{Sva96}A. Svane, Phys. Rev. B {\bf 53}, 4275 (1996).
\bibitem{Joh95}B. Johansson et al., Phys. Rev. Lett. {\bf 74}, 2335 (1995).
\bibitem{Bae91}Y. Baer, M. Grioni, and D. Malterre, Phys. Rev. B {\bf 44}, 9108-9109 (1991).
\bibitem{Raf09}M. Rafiee, and S. Jalali Asadabadi, Comp. Mat. Scie. {\bf 47}, 584-592 (2009).
\bibitem{Eas70}D.E. Eastman, Phys. Rev. B {\bf 2}, 1 (1970).
\bibitem{Lie70}L. Liebermann, J. Clinton, D.M. Edwards and J. Mathon, Phys. Rev. B {\bf 25}, 232 (1970).

\end{thebibliography}
\end{document}